\documentclass{article}  
\usepackage{bigsky2007}
\usepackage{graphicx}
\frompage{000} \topage{000}                                              
\def\rightmark{THGEM Application}
\def\leftmark{M. Gai {\em et al.}}
 
\begin{document}
 
\title{Toward Application of a Thick Gas Electron Multiplier (THGEM) Readout for a Dark Matter Detector
\footnote{This work is supported by the Yale-Weizmann Collaboration Program of the American Committee on Weizmann Institute of Science (ACWIS), New York.} } 
\authors{
{M. Gai$^1$, R. Alon$^2$, A. Breskin$^2$, M. Cortesi$^2$, D.N. McKinsey$^1$, J. Miyamoto$^2$, K. Ni$^1$, D.A.R. Rubin$^{1,2}$, and T. Wongjirad$^1$}\\[2.812mm]
{\normalsize
\hspace*{-8pt}$^1$ Department of Physics, Yale University, PO Box 208120, 
217 Prospect St, \\ New Haven, CT 06520-8120, USA.\\
\hspace*{-8pt}$^2$ Radiation Detection Physics Laboratory, The Weizmann Institute of Science
76100, Rehovot, Israel.
}}
 
\abstract{The {\bf Yale-Weizmann collaboration} aims to develop a low-radioactivity (low-background) cryogenic noble liquid detector  for Dark-Matter (DM) search in measurements to be performed deep underground as for example carried out by the XENON collaboration. A major issue is the background induced by natural radioactivity of present-detector components including the Photo Multiplier Tubes (PMT) made from glass with large U-Th content. We propose to use advanced Thick Gaseous Electron Multipliers (THGEM) recently developed at the Weizmann Institute of Science (WIS). These Òhole-multipliersÓ will measure in a two-phase (liquid/gas) Xe detector electrons extracted into the gas phase from both ionization in the liquid as well as scintillation-induced photoelectrons from a CsI photocathode immersed in LXe. We report on initial tests (in gas) of THGEM made out of Cirlex (Kapton) which is well known to have low Ra-Th content instead of the usual G10 material with high Ra-Th content.
}


\maketitle
\setcounter{page}{1}

\section{Introduction}\label{intro}

Astrophysical evidence on a variety of distance scales clearly shows that a large fraction (96\%) of the mass-energy density of the universe is unaccounted for. A large fraction (23\%) of the missing Òdark matterÓ must be non-baryonic, not constituted of the protons and neutrons that make up ÒordinaryÓ matter. A compelling explanation for the missing mass is the existence of Weakly Interacting Massive Particles (WIMPs). Their existence is well supported by particle-physics theories beyond the Standard Model, and thus the terrestrial discovery of WIMPs would have enormous impact on both astrophysics and particle physics \cite{Jun96}. WIMPs, if they exist, would occasionally interact with normal matter, be elastically scattered from nuclei and deposit a small amount of energy. With a large mass (e.g. the lightest super-symmetric candidate, the neutralino, is expected to have a mass of the order of 100 GeV), and moving at speeds of about 220 km/s relative to the Earth (the velocity of the Sun around the Milky Way), WIMPs would only deposit several tens of keV when scattered off nuclei. Because they would only interact weakly with ordinary matter, the rate at which WIMP particles would be scattered by ordinary matter is only 0.1 - 0.005 events per day per kg. In order to test the WIMP hypothesis, detectors must be built that are both low in background radioactivity and highly sensitive to low-energy nuclear recoils. New detectors that reach high sensitivity at low energy may be able to rigorously test the hypothesis that WIMP particles account for the missing mass of the universe.  
 
\section{The XENON Search for Dark Matter}\label{XENON}

The best prospects for the unambiguous identification of a WIMP signal lie in detectors that have negligible background competing with the dark-matter signal. This can be achieved principally by using nuclear recoil discrimination in order to veto competing electron recoil events (associated with gamma and beta backgrounds), effective neutron shielding, and through the operation of a large homogeneous self-shielded detector volume with 3-D position resolution. The 3-D position information can be used to select single-hit events characteristic of a WIMP interaction while rejecting multiple-hit events associated with background (e.g. neutron) events that propagate from the edge of the detector into the fiducial volume.

The XENON experiment \cite{XENON} in its initial configuration (XENON10) has thus far reached the best sensitivity for the direct detection of dark matter in the form of WIMPs, via their elastic scattering off Xe nuclei  \cite{XENON10}. In the full XENON detector a fiducial mass of 1000 kg, distributed in ten independent liquid-Xe time projection chambers (LXe-TPCs), will be used to probe the lowest predicted interaction cross section. The XENON10 experiment has thus far used 14 kg Xe \cite{XENON10}. The TPCs (see prototype scheme in Fig. 1) was  operated in dual (liquid/gas) phase, to allow a measurement of nuclear recoils down to 16 keV. They simultaneously detect, with Photo-Multiplier Tubes (PMTs), the primary scintillation signal in the liquid, and the ionization signal in the liquid through secondary scintillation in the gas induced by ionization electrons extracted from the LXe. The distinct ratio of primary to secondary scintillation (so called "S1/S2" ratio), respectively, measured by Òbottom PMTsÓ immersed in the LXe and Òtop PMTsÓ placed in the gas cell (Fig.1), for nuclear recoils from WIMPs (or neutrons), and for electron recoils from background, is the key to the event-by-event discrimination capability of the XENON TPC detector.

 \begin{figure}
 \begin{center}
 \includegraphics[height=2.5in]{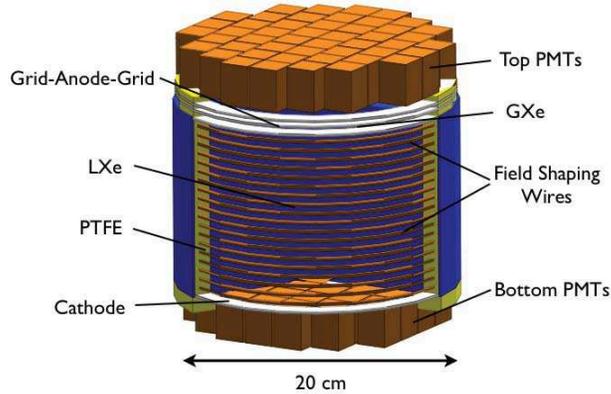}
 \end{center}
 \vspace{-1cm}
 \caption{\label{fig1} The XENON detector. Note the large number of PMTs; a large source of background.}
\end{figure}

One of the largest sources of background in the XENON (and most other underground) detector is the radioactivity contained in the glass housing of the vacuum PMTs. Glass contains a large amount (30 parts per billion) of alpha-emitters (U-Th) isotopes that produce a large ambient (beta and gamma ray) background as well as secondary neutrons produced by the interaction of the approximately 5 MeV alpha-particle with low mass materials (e.g. $^{10,11}$B and $^{13}C$ contained in the glass). For example for the PMTs of the XENON100 (100kg) detector, 2 neutrons per day produced in the PMTs of the detector were estimated. The sensitivity of the XENON (and most other underground) detectors searching for dark matter will be limited by this radioactivity and it is essential to remove the glass material, hence removing the PMT containing it. We set out to investigate alternative readout method that will replace the PMT used in the XENON and other underground experiments.

 \section{The Proposed Low-Background Readout}
 
 To circumvent the background from PMTs we aim to detect the charge produced by both ionization and scintillation signals of the LXe-TPC with a novel Thick Gas Electron Multiplier (THGEM) developed at the Weizmann Institute \cite{THGEM1,THGEM2,THGEM3} and recently tested in a modest size 
(10 cm diameter) board by the UConn-Weizmann-PTB collaboration \cite{Leo}. These Òhole-multipliersÓ will measure in a two-phase (liquid/gas) Xe detector both ionization electrons from the liquid as well as scintillation-induced photoelectrons from a CsI photocathode immersed in the LXe, extracted into and multiplied in the gas phase. In Fig. 2 we show a schematic diagram of the proposed detector. Our study is similar to that of Bondar {\em et al.} \cite{Bondar} except for the use THGEM in xenon and in our proposed detector the photo-cathode will be immersed in the liquid.

 \begin{figure}
 \begin{center}
 \includegraphics[height=2.2in]{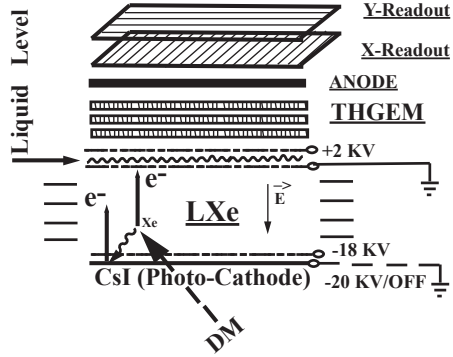}
 \end{center}
 \vspace{-1cm}
\caption{\label{fig.2}A schematic diagram of the proposed Dark Matter detector.}
\end{figure}

 \subsection{The Cirlex-THGEM Electron Multiplier}
 
The THGEM we investigated were made of Cirlex (polyimide or kapton) that was tested by the XENON collaboration to have about 30-fold smaller radioactivity than glass. We purchased a number of 13Óx26Ó double sided copper clad Cirlex boards 0.4 mm thick from Fralock \cite{Cirlex} and fabricated \cite{Israel} fifteen samples of Cirlex-THGEM thick boards as designed at the Weizmann Institute and shown in Fig. 3. One particularly good aspect of the boards produced by Fralock is that the copper cladding is adhered to the board using a proprietary method that does not involve additional material or glue. Hence the board includes only polyimide (kapton) and copper. 

 \begin{figure}
 \begin{center}
 \includegraphics[height=1.3in]{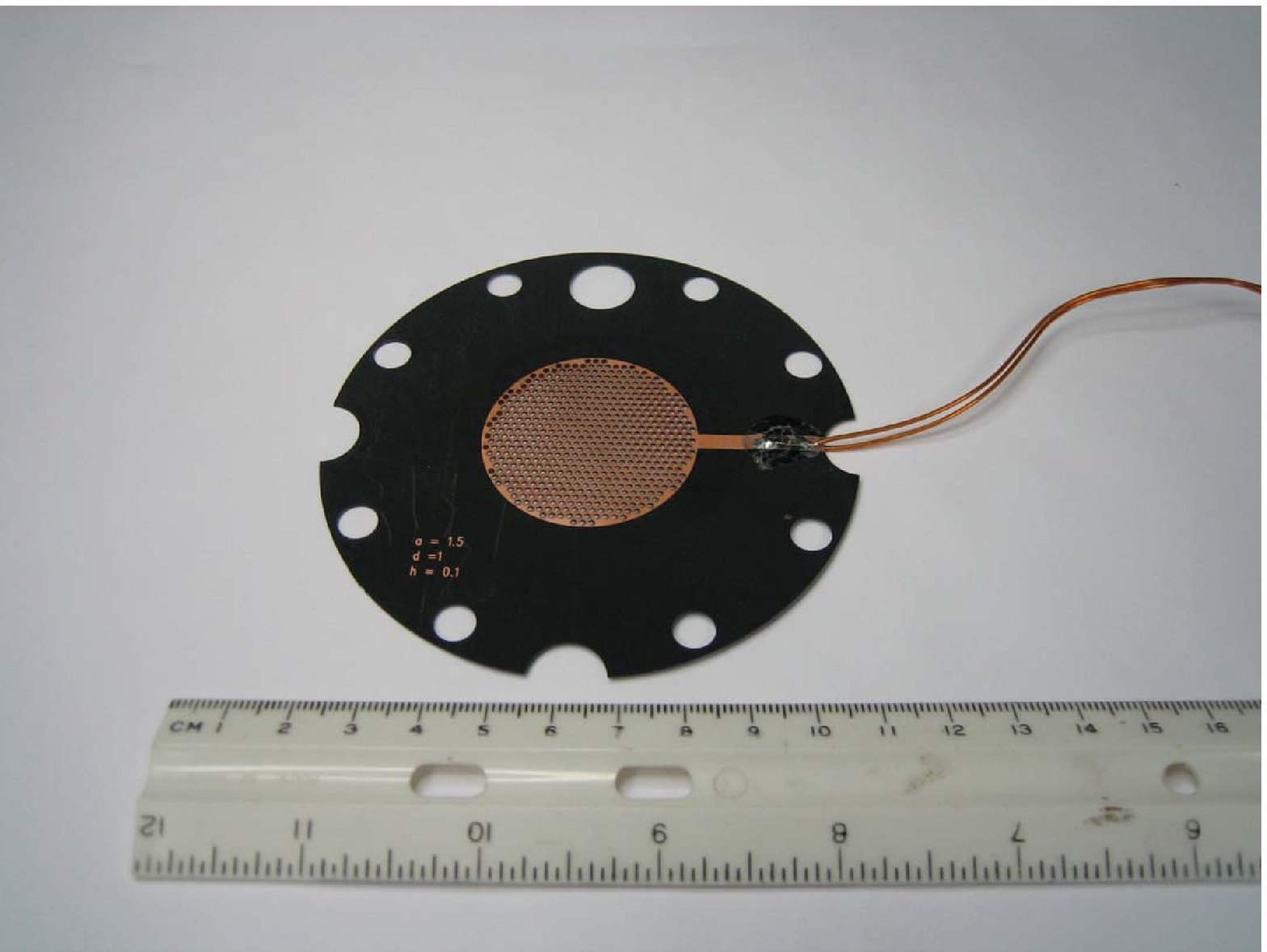} \includegraphics[height=1.3in]{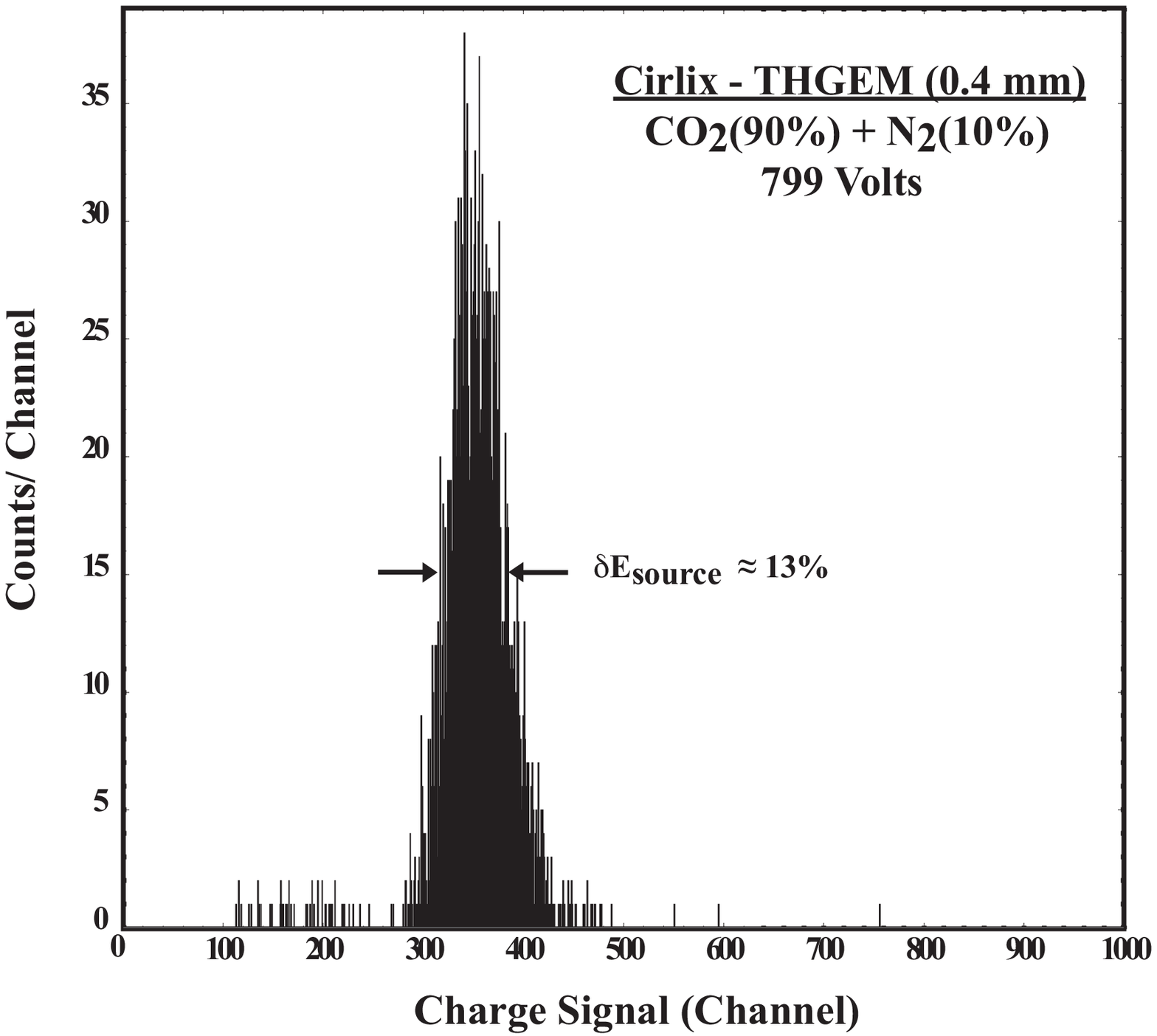}
 \end{center}
 \vspace{-0.7cm}
\caption{\label{fig3}A picture of the Cirlex-THGEM board used in this study and the obtained charge signal.}
\end{figure}

The Cirlex THGEM boards were tested at the LNS at Avery Point and the obtained signal is shown in Fig. 3. Note that the peak width is close to the 13\% energy distribution of the non-spectroscopic sealed $^{241}Am$ alpha-particle source used in this study. In Fig. 4 we show a comparison of the electron multiplication (gain) obtained with single Cirlex-THGEM and G10-THGEM in a gas mixture and pure argon. The Cirlex-THGEM boards are observed to produce comparable gains. In the same figure we show the results of  measurements of double G10-THGEM performed at the Weizmann institute with argon gas at higher pressures.

 \begin{figure}
 \begin{center}
 \includegraphics[height=2in]{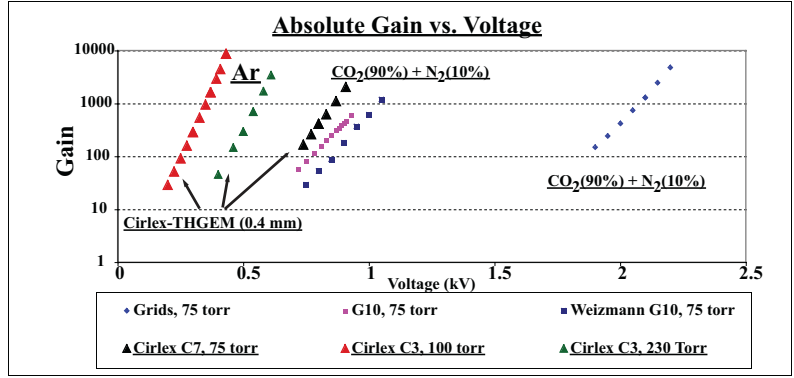} \includegraphics[height=2in]{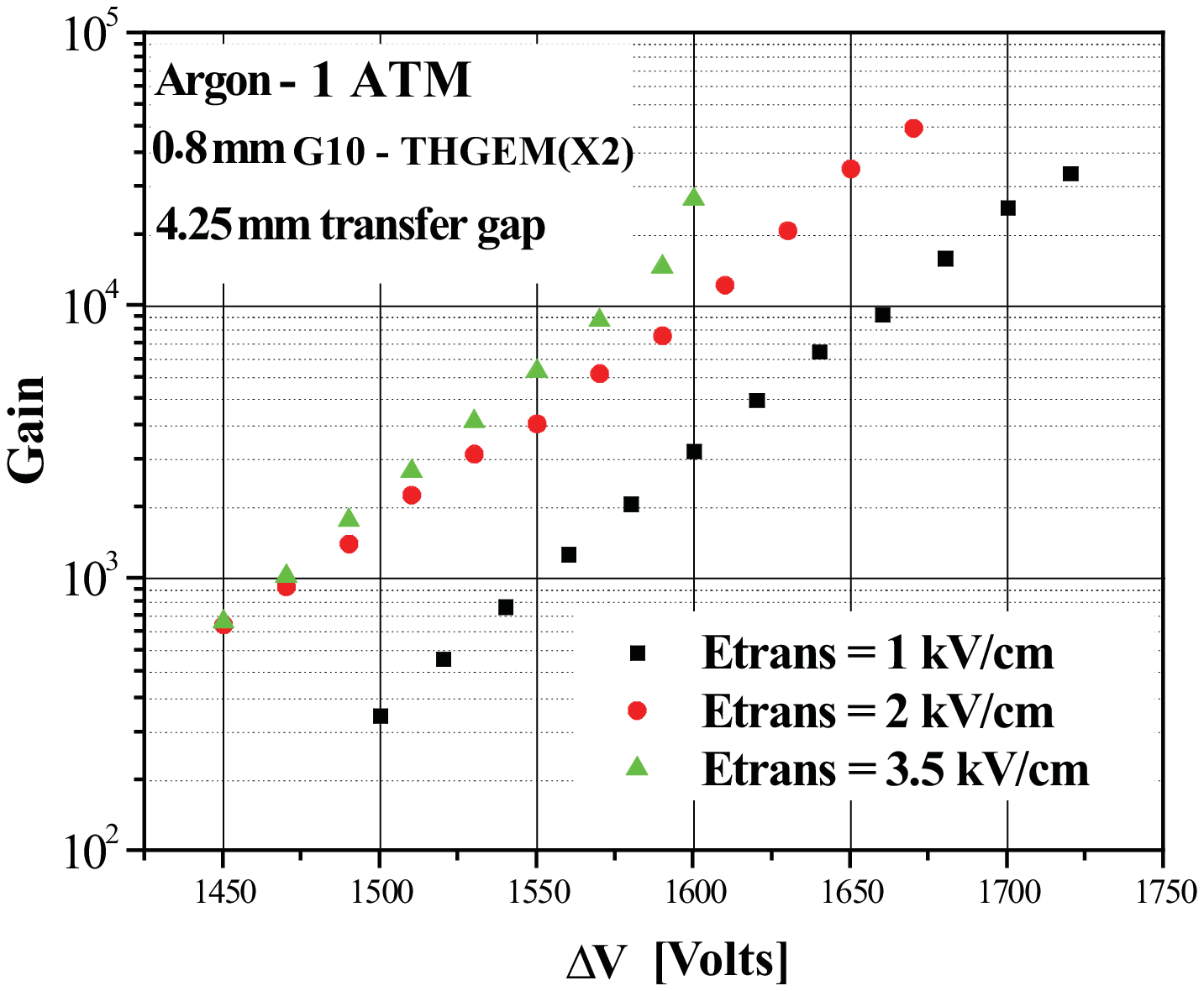}
 \end{center}
 \vspace{-0.7cm}
\caption{\label{fig4}Gain curves obtained with single Cirlex-THGEM and other charge multipliers (top figure) and with with double G10-THGEM at higher pressures (bottom figure).}
\end{figure}

\begin{figure}
 \begin{center}
\includegraphics[height=1.5in]{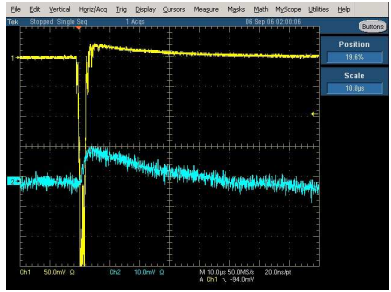}
 \end{center}
 \vspace{-0.7cm}
\caption{\label{fig5}Light signal (top trace) and amplified charge signal (bottom trace) obtained from the xenon cell operated at room temperatures.}
\end{figure}

Measurement in the Yale xenon cell were performed with the Cirlex-THGEM immersed in xenon gas at room temperature. The prompt light emission was detected using the PMT placed at the bottom of the xenon cell and the charge signal was measured from the top of the Cirlex-THGEM. The obtained light and charge signals are shown in Fig. 5.

\section{Conclusions}
THGEM boards made out of Cirlex were fabricated and tested in gas phase at the LNS at Avery Point, G10 THGEM boards were tested at Weizmann, and initial tests in a xenon cell were carried out at Yale. These tests indicate Cirlex-THGEMs are useful for the proposed low background readout. The installation of a very accurate ($\pm 10 \mu m$) board drilling machine at Brookhaven National Lab (BNL) \cite{BNL} will allow for a continuation and expansion of this study.


\begin{thebibliography}{9}
\bibitem{BNL} A high precision circuit board drilling machine was recently purchased 
from Schmollen Machinen, Germany, by the Instrumentation Division of Brookhaven National Lab.
\end{thebibliography}
\end{document}